\documentclass[useAMS,usenatbib,usegraphicx]{mn2e}

\usepackage{journals}
\usepackage{multirow}
\usepackage{url}
\usepackage{threeparttable}
\usepackage{graphicx}

\usepackage{color}
\usepackage{ulem}

\title[Period--luminosity relation for LMXBs in the NIR]{Period--luminosity relation for persistent LMXBs in the near-infrared}
\author[Revnivtsev et al.]{Mikhail G. Revnivtsev$^{1}$\thanks{E-mail: revnivtsev@iki.rssi.ru}, Ivan Yu. Zolotukhin $^{2,3,4}$ and Alexander V. Meshcheryakov $^{1}$\\
$^{1}$Space Research Institute, Russian Academy of Sciences, Profsoyuznaya 84/32, 117997 Moscow, Russia\\
$^{2}$Observatoire de Paris-Meudon, LERMA, UMR~8112, 61 Av. de l'Observatoire, 75014 Paris, France\\
$^{3}$Sternberg Astronomical Institute, Moscow State University, Universitetskij pr., 13, 119992, Moscow, Russia\\
$^{4}$Observatoire de Paris, VO-Paris Data Centre, 61 Av. de l'Observatoire, 75014 Paris, France\\
\\
}
\begin{document}

%\date{Accepted 2011 August XX. Received 2011 July XX; in original form 2011 July 21}

\pagerange{\pageref{firstpage}--\pageref{lastpage}} \pubyear{2011}

\maketitle

\label{firstpage}

\begin{abstract}
We study relations between the X-ray luminosity, orbital period and absolute near-infrared magnitude of persistent low-mass X-ray binaries (LMXBs). We show that often optical and near-infrared spectral energy distribution of LMXBs can be adequately described by a simple model of an accretion disc and a secondary star reprocessing X-ray emission of a central compact object. This gives us an evidence that using an X-ray luminosity and an absolute infrared magnitude of a persistent LMXB one can make reliable estimate of its orbital period. Using a sample of well-known LMXBs, we have constructed a correlation of $L_{\rm X}, P_{\rm orb}$ and $M_{K}$ values which can be approximated by a straight line with the RMS scatter at the level of $\sim$0.3~mag. Such a correlation, being to some extent an analogous to the correlation, found by \citet{paradijs94}, might be helpful for future population studies especially in the light of forthcoming surveys of the Galaxy in X-ray and infrared spectral domains.
\end{abstract}


\section{Introduction}

Low-mass X-ray binaries are binary systems with compact objects (neutron stars or black holes) accreting matter from a low-mass secondary companion. First such systems were discovered about 50 years ago \citep{giacconi62} and presently more than a hundred of them are known in our Galaxy \citep[e.g.][]{liu07} and hundreds of them in external galaxies \cite[see e.g.][]{fabbiano06,evans10}. Population of low-mass X-ray binaries provides us with tools to study different physical mechanisms, that define the population features (e.g. the energy transfer efficiency during the common envelope stage of the binary, the gravitational wave emission influence, the average magnetic field strength in old neutron stars), and also some properties of a host galaxy (e.g. the age of its stellar population). In order to reach these goals one needs: (1) to better understand the laws of formation and evolution of LMXBs, and (2) to gather a clear sample of LMXBs with known parameters (orbital periods, X-ray luminosities, etc).

One of the major obstacles with the latter item is the lack of measurements of orbital periods for the majority of known LMXBs. This is mainly caused by the fact that the orbital modulation of the X-ray emission is rarely observed due to geometrical reasons (small companion star covers insignificant fraction of the sky for the compact object, therefore very few known LMXBs demonstrate eclipses and/or dips), whereas the detection of the optical or infrared (IR) modulation of LMXBs emission is challenging due to extreme faintness of these objects.

However, it is sometimes possible to estimate the LMXB orbital periods indirectly, making use of the fact that their optical and IR emission originates mainly from reprocessing of the central X-ray emission by the outer accretion disc \cite[see e.g.][]{mcclintock79,paradijs94}. The efficiency of this approach was demonstrated by \cite{paradijs94}, who showed that the absolute optical magnitude of LMXBs had very clear correlation with their X-ray luminosity and orbital period. It should be kept in mind, however, that this method is physically justified for {\it persistent} LMXBs, which possess stationary accretion flow with a constant incident X-ray  flux. A transient-like behaviour might lead to the large variations of X-ray flux or/and the accretion flow structure and thus to the deviations from any possible correlation. One might also expect significant deviations if the central X-ray emission of the source is hidden from an observer due to the very large inclination angle of the binary system (so called Accretion Disc Corona/ADC sources like 4U1822-37).

Unfortunately, the majority of LMXBs in the Galaxy (either transient or persistent) reside in the Galactic plane  and thus suffer from severe dust extinction reaching $A_V\sim10-30$~mag making them completely unobservable in the optical domain, thus shifting observational methods to longer wavelengths. Due to the fact that reliable measurements of LMXBs absolute magnitude in the infrared spectral bands can be done for much more sources, it is then reasonable to calibrate diagnostics similar to that of \cite{paradijs94} in the IR, in particular, in the $K$ spectral band. 

However, spectral energy distribution (SED) of sources at these wavelengths
might be contaminated by optically thin synchrotron emission, sometimes observed
in the far end of NIR up to radio range in persistent \citep{migliari10} and
especially often in transient LMXBs \cite[see
e.g.][]{corbel02,russell06,russell07,shahbaz08,russell08}. This issue is to be
carefully explored before we are able to construct reliable period--luminosity
relation for the $K$ band.

In this paper we show that the general shape of SED of {\sl persistent} LMXBs in the optical--NIR spectral range can be reasonably well described in the framework of a model, where the optical--NIR emission comes from the reprocessing of the central X-ray flux in the accretion disc and the secondary star. This gives us an evidence that there is a physical connection between the main properties of the binary and its infrared luminosity. We present the correlation of an absolute magnitude of LMXBs with their X-ray luminosity and orbital period similar to the one found in the $V$ band by \cite{paradijs94}. In this work we consider only persistent LMXBs to be sure that their accretion disc structure does not change much at different epochs. We consider only LMXBs which harbour neutron stars as a compact object in order to reduce the uncertainties related with the mass of the compact object.

\section{LMXBs spectral energy distribution in the optical and NIR}

In order to check the validity of our approach to interpretation of the LMXB
luminosity in optical and NIR spectral domains, we have studied broad band
spectral energy distribution of the three binary systems, Cyg X-2, Sco X-1,
and 4U0614+091, which can be considered as illustrative. These three systems
cover wide range of orbital periods, from 0.81 to 236.27~h, and have more
or less reliable distance, interstellar extinction and in two
cases (Cyg X-2 and Sco X-1) inclination estimates. Their parameters, adopted
in this study for spectral modelling, are presented in Table~\ref{params}.

It is important to keep in mind that in some cases a considerable
contribution from an optically thin emission can be seen at different states of
X-ray binaries \cite[see e.g.][]{russell06}. In particular, detection of
polarized infrared emission of some X-ray binaries, exceeding levels expected
from interstellar scattering, hints at synchrotron origin of some fraction of
their infrared flux \cite[see e.g.][]{shahbaz08}.

Therefore, in order to test the applicability of our model that accounts for the optically thick emission, we have intentionally selected X-ray binaries at different spectral states, namely high/soft (Sco X-1 and Cyg X-2) and low/hard (4U0614+091) state.

For calculation of spectral energy distribution of LMXBs we adopt a simple model \cite[see e.g.][]{tjemkes86,obrien02,mescheryakov11b} when all surface elements in a binary system (which comprises an accretion disc and a secondary star) emit black body radiation with the local temperature defined by X-ray irradiation and internal heating. Shape of the star was assumed to be of Roche lobe geometry.

Temperature of the illuminated side of the star
  was calculated taking into account heating due to incident X-ray flux:
 \begin{equation}
 T_{\star}^4 = T_{\star,0}^4 +
  {{\eta_\star L_{\rm X} \cos\theta_\star}\over{4\pi\sigma d^2}},
\end{equation}
where $L_{\rm X}$ -- luminosity of the central X-ray source, $\theta_\star$ --
angle between X-ray source direction and surface normal, $d$ -- distance from
X-ray source to the surface element, $\sigma$ -- Stefan-Boltzmann constant,
$\eta_\star$ -- fraction of reprocessed X-ray emission. Temperature of the part of the star not illuminated by a compact object was taken following \cite{dejong96} as $T_{\star,0}=5800 (L/L_\odot)^{1/4}(R/R_\odot)^{-1/2}$~K and mass--luminosity and mass--radius relation from \cite{tout96}. For the fraction of X-ray emission reprocessed and reradiated from the surface of the
  secondary star we assume $\eta_\star=0.6$ \cite[see e.g.][]{london81}.

  For the effective temperature of the disc surface element we used the
  common relation, which includes viscous heating and heating by X-ray
  irradiation in the outer parts of geometrically thin accretion disc:
 \begin{equation}
   T_{\rm d}^4 = {{3 G M_1 L_{\rm X}}\over{8\pi \epsilon c^2\sigma R^3}} +
   {{\eta_{\rm d} L_{\rm X}}\over{4\pi\sigma R^2}} \left(\frac{H}{R}\right)_{\rm out}(n-1).
\label{eq:Td} 
\end{equation} 
We use $M_1=1.4\,{\rm M_\odot}$ and $\epsilon=0.1$, -- mass of the compact object
and accretion efficiency, respectively, for NS LMXB.

The fraction of the flux, intercepted by the accretion disc, was calculated assuming a shape of the intercepting surface as disk with height being a function of its radius: $H \propto R^n$, we assume $n=9/7$ \citep{vrtilek90}. We note here that intercepting surface is not necessarily the optically thick accretion disc itself, it might be the corona that intercepts the X-ray flux and redirects it to the optically thick accretion disc, see e.g. \citet{jimenez02,mescheryakov11a}. We assume ratio of hight of the intercepting surface to its radius at the outer edge of the disc $(H/R)_{\rm out}=0.1$.  Parameter $\eta_{\rm d}$ -- the effective fraction of
emission, reprocessed in the accretion disc--corona system (ratio of X-ray
flux thermalised in optically thick disc to the incident X-ray flux) is not
yet accurately known, different authors estimate its value from less
than 0.1 \citep{dejong96} up to $\sim$0.5 \citep{vrtilek91}. In
our work we adopt the value $\eta_{\rm d}=0.25$.

The accretion disc radius around the compact object was taken to be equal to the tidal radius as estimated by \cite{p77}. More detailed description of our model can be found in \cite{mescheryakov11b}. We would like to note here that our ability to predict the
absolute magnitude of LMXBs with the accuracy better than 0.5--0.8~mag (i.e.
within a factor of two in the flux) mostly depends on the accuracy of our
knowledge of the binary system inclination and the fraction of reprocessed
X-ray emission (for fixed disc radius and shape of the X-rays intercepting
surface).

\begin{table}
\caption{Parameters of the three binary systems, for which we have calculated optical--NIR SEDs. Orbital periods were adopted from \citet{ritter03}. Cyg X-2 and Sco X-1, being members of the so called Z-sources, were assumed to have the Eddington luminosity, whereas the luminosity of 4U0614+091 is taken from \citet{revnivtsev11}. See the text for other parameters, common for these systems.}
\label{params}
\begin{tabular}{l|c|c|c}
		& Cyg X-2	& Sco X-1 & 4U0614+091\\
\hline
Distance(kpc)&$11.6\pm0.3^1$&$2.8\pm0.3^2$&3.2$^3$\\
$P_{\rm orb}$&236.27&18.94&0.81\\
$M_1$&1.71$^{4}$&1.4	&1.4	\\
$q$&0.34$^{4}$&0.3$^{5}$&0.04$^{6}$	\\
$\log L_{\rm x}$(erg/sec)&38.3&38.3&36.5\\
Inclination&62.5$^7$&38$^{5}$&60$^{8}$\\
$A_V$&1.34$^9$&0.91$^10$&2$^{11}$\\
\end{tabular}
\begin{tablenotes}
\item (1) -- \cite{smale98}, (2) -- \cite{bradshaw99}, (3) - \cite{kuulkers10}, (4) -- \cite{casares10}, (5) -- \cite{steeghs02}, (6) -- \cite{werner06}, (7) -- \cite{orosz99}, (8) -- inclination is unknown, fixed at given value, (9) -- \cite{mcclintock84}, (10) -- \cite{vrtilek91}, (11) -- \cite{migliari10}
\end{tablenotes}
\end{table}

\section{Data}

 In this work we consider only {\sl persistent} LMXBs, which means that their
optical, IR and X-ray fluxes do not vary by large factors. This is essential for
present work, because we use here measurements of brightness of sources in
optical, IR and X-ray energy bands, obtained not simultaneously.

Distribution of X-ray fluxes of bright (with fluxes more than $\sim$50 mCrab, i.e. which have enough statistics to be detected by All-Sky Monitor (ASM) of {\it RXTE} observatory in all single measurements) persistent LMXBs were analysed by \cite{postnov05}. It was shown there that the typical long-term root-mean-square flux deviations do not exceed $\sim$20--30~per~cent of their mean values.
Long-term variations of optical emission of persistent LMXBs were studied in a number of works, in particular for sources Sco X-1 \citep{mcnamara05}, Cyg X-2 \citep{obrien04}, and 4U0614+091 \citep{hakala11} considered in our paper. It was demonstrated that the long-term variations of their apparent optical magnitudes do not typically exceed $\sim$0.1--0.3~mag (i.e. $\sim$10--30~per~cent flux variation). 

Based on these evidences we assume additional $\sim$10--30~per~cent  systematic
flux uncertainty due to non-simultaneous nature of the measurements we use while
analysing X-ray and optical luminosity correlations.

To construct a broad band spectral energy distributions of our test objects we have used their multi-colour photometry from the literature and also new publicly available data from {\it Spitzer} (instruments IRAC \citep{fazio04} and MIPS \citep{rieke04}) and {\it WISE} \citep{wright10} space observatories.

Mid-infrared photometric measurements were obtained from the analysis of {\it Spitzer} basic calibrated data images. For each source we have determined the flux by summing counts in circular aperture with the radius of 6~arcsec (for {\it Spitzer}/MIPS at 24~$\mu$m we have used 9 arcsec aperture due to larger FWHM of the instrument PSF at these wavelengths) centered at the object and subtracting the background counts, collected in an appropriate field nearby.

Another set of photometric measurements was extracted from preliminary data release of the {\it WISE} satellite \citep{wright10}. Conversion of magnitudes into physical fluxes was achieved by using the coefficients, presented in \cite{cutri11}.

All photometric measurements were corrected for the interstellar extinction assuming the \cite{rieke85} reddening law.

 Data at these long wavelengths are important for additional check for the
presence of optically thin synchrotron emission, sometimes seen in SEDs of LMXBs
\cite[see e.g.][]{migliari10}.

\subsection{Cyg X-2}

Photometric measurements of Cyg X-2 used in this work are given in Table~\ref{cygx2_table}. Model of the broadband SED, constructed assuming the above mentioned parameters is shown by solid curve. The model is not fitted to the data, but presented as it is, making use of the parameters, described in the text and in Table~\ref{params}. Accurate position of the modelled SED with respect to Y-axis depends on the source distance, the fraction of emission reprocessed in thermal disc and the binary system inclination. 

\begin{table}
\caption{Photometric measurements of Cyg X-2. We have assumed its interstellar extinction to be $A_{V}=1.34$.}
\label{cygx2_table}
\begin{center}
\begin{tabular}{lccc}
Filter&Vega mag&$A_\lambda$&Vega mag corr.\\
\hline
$U$	& 15.0$^1$& 2.05	&	12.9\\
$B$	& 15.3$^1$& 1.77	&	13.5\\
$V$	& 14.8$^1$&	1.34	&	13.5\\
$R$	& 14.0$^2$&	1.00	&	13.0\\
$J$	& 13.39$^3$&0.38	&	13.01\\
$H$	& 13.15$^3$&0.24	&	12.92\\
$K$	& 13.05$^3$&0.15	&	12.90\\
\end{tabular}
\vspace{5mm}
\begin{tabular}{lc}\\
\multicolumn{2}{c}{\it{Spitzer}}\\
\hline
$\lambda$, $\mu$m&Flux, mJy\\
4.5&1.4\\
8.0&0.8\\
24& 8.5\\
\end{tabular}
\end{center}
\begin{tablenotes}
\item (1) -- \cite{orosz99}, (2) -- USNO B1, (3) - 2MASS
\end{tablenotes}
\end{table}

\begin{figure}
\includegraphics[height=\columnwidth,bb=28 170 570 710,clip]{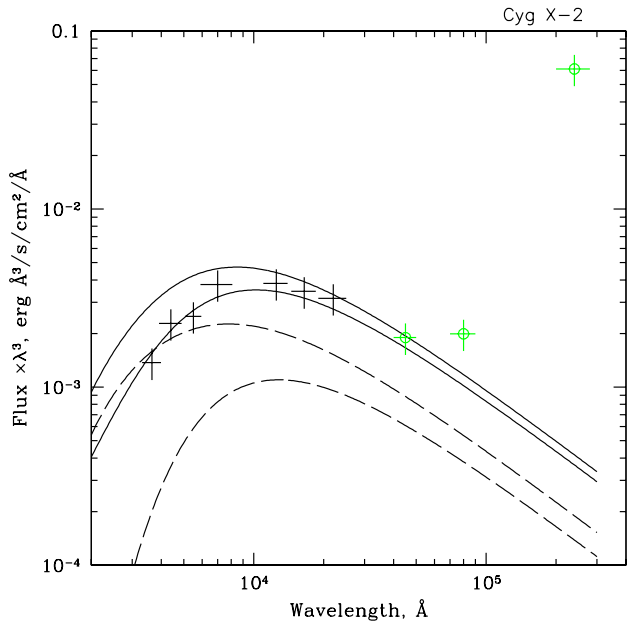}
\caption{Extinction-corrected spectral energy distribution of Cyg X-2 in the
optical and IR with a simple model of its emission (see text). The model is not
fitted to the data, but presented as it is, making use of the parameters,
described in the text and in Tab.~\ref{params}. Dashed line represents
contribution of the secondary star. Two sets of curves represent SEDs during
maximum and minimum contribution of secondary star to the total brightness of
the binary. Due to possible time variations of the optical/NIR emission of the
source we have assumed 0.2~mag uncertainty of all photometric measurements. Open
circles denote measurements obtained using {\it Spitzer} data.}
\label{cygx2}
\end{figure}

It is clear from Fig.~\ref{cygx2} that the adopted model adequately describes
the shape of the broad band spectral energy distribution of the source, except
for its long wavelength range (8 and 24~$\mu$m). We make the following
conclusions from this comparison: 1) the model well describes the SED of Cyg X-2
up to $\sim$4~$\mu$m, and 2) at longer wavelengths we observe the indication of
some additional emission component which is likely the optically thin
synchrotron emission of non-thermal electrons. Polarization measurements,
indicating that at these wavelengths the source might have a contribution from
synchrotron component were presented by \cite{shahbaz08}.

\subsection{Sco X-1}

A set of Sco X-1 photometric measurements in the optical and infrared, obtained at different epochs with different instruments, is presented in Table~\ref{scox1_table}. It is clearly seen that Sco X-1 exhibits significant variations of its spectral energy distribution at long wavelength end, while in the optical the changes are inessential \cite[see e.g.][]{mcnamara05}.
It is not surprising because the source is known to be strongly variable at radio wavelengths \cite[e.g.][]{pandey07}, which means that it sometimes displays a spectral component emerging from non-thermal population of electrons. In addition to that it was shown that in the NIR Sco X-1 sometimes exhibits polarization, exceeding the one induced by the interstellar dust scattering \citep{shahbaz08,russell08}, which can also be attributed to the emission caused by non-thermal electrons. Finally, yet another spectral component (different from the optically thick emission, considered by our model) can be found in the presence of hard X-ray tails, often observed from Sco X-1 \cite[e.g.][]{damico01,paizis06}, and likely to be related to the emission of energetic non-thermal electrons in the accretion flow \cite[see e.g.][]{migliari07}.

\begin{figure}
\includegraphics[height=\columnwidth,bb=28 170 570 710,clip]{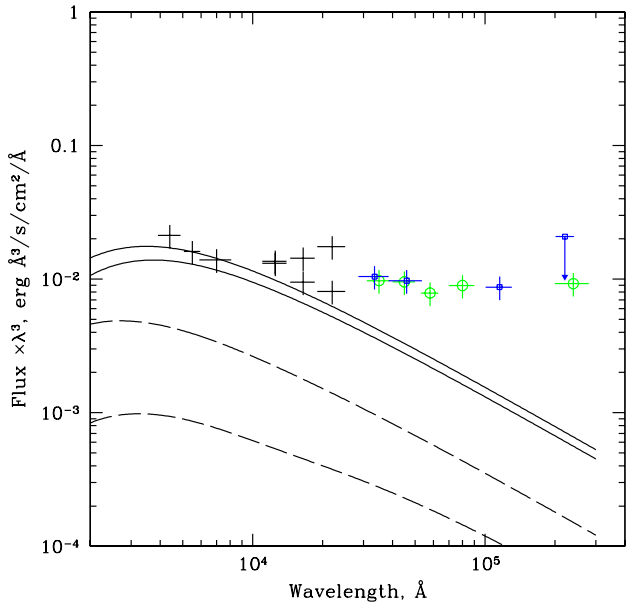}
\caption{The same as Fig.~\ref{cygx2}, but for Sco X-1. Open circles denote measurements obtained using {\it Spitzer} data, open squares -- measurements from {\it WISE} data.}
\label{scox1}
\end{figure}

However, in spite of these complications we note that one set of NIR measurements (namely those of \citealt{willis80}) does show the Rayleigh--Jeans type SED in NIR suggesting that the contribution of an additional component attributable to non-thermal electrons at that epoch was small or negligible. We will therefore use this $K$ magnitude estimate in our subsequent work.
Importantly, the $J-K$ colour can be used as an indicator of a presumably synchrotron component in the SED. In particular, our model predicts $J-K\sim-0.15$~mag for the source (see below), while the brightest NIR measurements reach $J-K=0.6$~mag. The faintest NIR estimates give $J-K=-0.19$~mag, which is reasonably compatible with the predictions of our model.

\begin{table}
\caption{Photometric measurements of Sco X-1. We have assumed its interstellar extinction to be $A_V=0.91$.}
\label{scox1_table}
\begin{center}
\begin{tabular}{lccc}
Filter&Vega mag&$A_{\lambda}$&Vega mag corr.\\
\hline
$B$     &12.38$^{1}$ 	&1.78&11.04   \\
$V$     &12.40$^{2}$	&0.91&11.49	\\
$R$     &12.3$^{3}$	&0.68&11.62    \\
$J$    &11.90$^{4}$	&0.25&11.65   \\
$H$    &11.54$^{4}$	&0.15&11.39     \\
$K$    &11.15$^{4}$	&0.10&11.05     \\
$J$    &11.94$^{5}$	&0.25&11.69    \\
$H$    &11.99$^{5}$	&0.15&11.84     \\
$K$    &11.98$^{5}$	&0.10&11.88    \\
\end{tabular}
\begin{tabular}{lc}
\multicolumn{2}{c}{{\it Spitzer}}\\
\hline
$\lambda$, $\mu$m& Flux, mJy \\
3.5      & 9.3\\
4.5      & 7.1\\
5.8      & 4.5\\
8.0      & 3.7\\
24       & 1.3\\
\multicolumn{2}{c}{\it{WISE}}\\
\hline
  3.35    &   10.4\\
  4.60    &   7.0\\
  11.5    &   2.5\\
  22.0    &   $<$3.0\\
\end{tabular}
\end{center}
\begin{tablenotes}
\item (1) -- \cite{mcnamara05}, (2) -- \cite{mcnamara03}, (3) -- USNO B1, (4) -- 2MASS, (5) -- \cite{willis80}
\end{tablenotes}
\end{table}
\subsection{4U0614+091}

\begin{figure}
\includegraphics[height=\columnwidth,bb=28 170 570 710,clip]{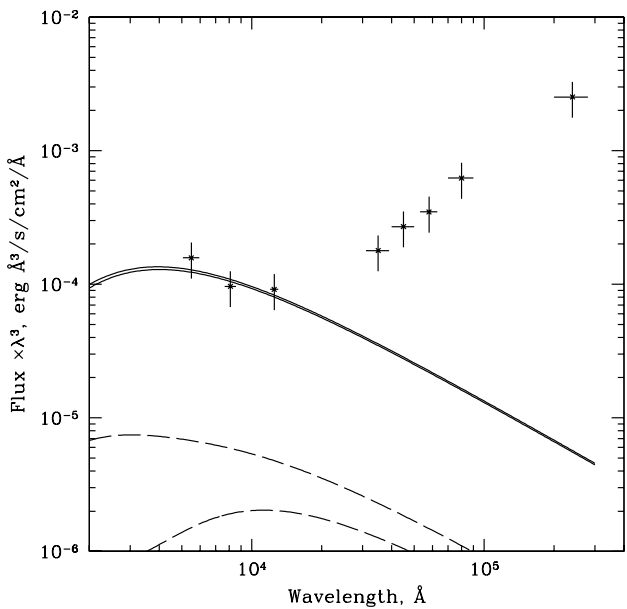}
\caption{The same as Fig.~\ref{cygx2}, but for 4U0614+091.}
\label{0614}
\end{figure}

We have adopted photometric measurements of 4U0614+091 from \cite{migliari10}. It is clearly seen (and it was mentioned in \citealt{migliari10}) that the source demonstrates optically thin component at mid-infrared wavelengths and its contribution to $K$ band can be comparable to that of the optically thick mechanism, considered by our model. For that reason we have calculated $K$ magnitude of 4U0614+091 from $J$ band measurement by \cite{migliari10}, using the model of optically thick emission (see Fig.~\ref{0614}), and used this value in the subsequent work.

\section{$M_K-\Sigma_{K}$ relation of persistent LMXBs}

\subsection{Model}

 Broad band spectral energy distributions, shown in the previous section,
provide us support that we adequately understand physical processes responsible
for the observational appearance of LMXBs in the optical and NIR spectral
domains.

According to this model the absolute optical and NIR magnitude of LMXBs
depend both on X-ray luminosity of the central source and on the size of the
binary system. \cite{paradijs94} showed that the luminosity of
  the system in optical $V$ band scales approximately as a square root of
its X-ray luminosity and as a power of 2/3 of its orbital period.
 Note that these indexes are theoretically
  expected for $V$ band photometry if temperatures of matter
  that gives the main contribution to optical emission lie in particular
  range of temperatures ($T\approx$ 10000--30000~K). For these 
  temperatures the surface brightness of a black body
emitter scales as $S_{V}\propto T^{\simeq2}$. For our case
of the NIR $K$ band the scalings should be different.

 In persistent LMXBs we expect that effective temperature of accretion
  disc is everywhere above hydrogen recombination limit. The effective
  temperature at which a hot disc becomes thermally unstable is, according
  to \cite{dubus01}, given by
\begin{equation}
T_{\rm H} = 7200 \alpha ^{-0.002} \left({{M_1}\over{M_\odot}}\right)^{0.03}
\left({{R}\over{10^{10}\,{\rm cm}}}\right)^{-0.08} {\rm K}
\label{eq:TH}
\end{equation}

Here $\alpha$ is the standard Shakura-Sunyaev viscosity parameter.

For the temperature range $T>T_{\rm H}$ the $K$ spectral band ($\lambda\approx 2.2$~$\mu$m)
lies almost at the Rayleigh-Jeans part of the disc blackbody spectrum,
therefore the disc surface brightness should scale as $S_{K}\propto
T^{\simeq1}$. As the outer disc temperature dominated by irradiation
  scales as $T_{\rm{out}}\propto L_{\rm X}^{1/4}R_{\rm out}^{-1/2}$ (see
Eq.~\ref{eq:Td}), having in mind $R_{\rm out} \propto a$ and
  $a\propto P^{2/3}$ (from Kepler's law) as well as $L_K \propto S_K R_{\rm out}^2$, the total IR luminosity of the accretion disc should be proportional to $L_{K}\propto L_{\rm X}^{0.25} P^{1}$.

  It is worth noting, that the above relations are
  simplistic, e.g. the noticeable viscous heating in the
    disc or contribution from the secondary star are expected to modify
    them. Therefore in order to understand more correct dependencies
  between the absolute NIR magnitude of LMXBs, their X-ray luminosity and
  accretion disc size (i.e. orbital period), we have calculated a set of LMXB models for the range of
  orbital periods of 2--750~h and X-ray luminosities of
  $1\times10^{35}$--$4\times 10^{38}$~erg~s$^{-1}$.

While varying the orbital period, we have assumed that the secondary star fills
its Roche lobe. We assumed that at orbital periods less then $\sim$6~h
($M<0.6\,{\rm M_\odot}$) the secondary star is not evolved and obey mass--radius
relation for a main sequence star \cite{tout96}. For the orbital periods larger
than $\sim$6~h (i.e. for the giant companions) we have fixed the companion
masses at $0.6\,{\rm M_\odot}$ value. This assumption in fact does not
influence our results strongly, and is supported by the cases when masses of the
donor giants are known, because they do not exceed this value, e.g. Sco X-1
$M_2\sim0.4\,{\rm M_\odot}$ \citep{steeghs02}, Cyg X-2 $M_2\sim0.6\,{\rm
M_\odot}$ \citep{orosz99}.

\begin{figure}
\includegraphics[height=\columnwidth,bb=28 170 570 710,clip]{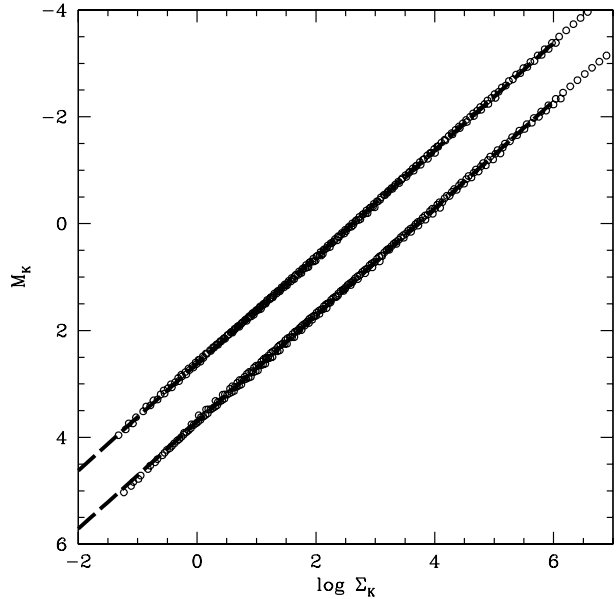}
\caption{Absolute NIR magnitudes of a modelled binaries versus their orbital period and X-ray luminosity, combined into $\Sigma_{K}$ quantity. Upper points represent absolute brightness for binaries with orbital plane inclination $i=0^\circ$, in this case $\Sigma_{K}=(L_{\rm X}/L_{\rm Edd})^{0.29} P({\rm h})^{0.92}$. Lower points show absolute brightness for binaries with inclination $i=70^\circ$, in this case $\Sigma_{K}=(L_{\rm X}/L_{\rm Edd})^{0.28} P({\rm h})^{0.93}$. Dashed lines show the best linear fit to points.}
\label{sigma_mk_model}
\end{figure}

After we have fixed the radius of the accretion disc (at the value of the tidal radius) and the shape of the disc (i.e. power law index $n$ in the law $H\propto R^{n}$), the main parameter, which can significantly shift the correlation up or down on the absolute magnitude axis, is 
the inclination angle of the binary system (this effect is illustrated in Fig.~\ref{sigma_mk_model}). Additional shift can be caused by different value of the fraction of emission reprocessed in the accretion disc--corona system. We have assumed it to be 0.25 above.

Among all models, covering the above-mentioned parameter space we have accepted only those, which have the disc temperature at outer radius not lower than $T_{\rm H}\approx6500$~K (see
  Eq.~\ref{eq:TH}). This selection was applied because we are interested only in persistent LMXBs, while LMXBs with lower outer disc temperatures should be subject to disc thermal-viscous instability (see e.g. \citealt{lasota01} for review).

The obtained absolute NIR magnitudes were fitted as a linear combination of $\log P$ and $\log L_{\rm X}$ parameters. We have minimized the root mean square deviations from the modelled $M_{\rm K}$ values. 
The best-fitting relation for LMXBs with the inclination of $i=0^\circ$ is $M_{ K{\rm,model,i=0}}=2.62-0.73 \log (L_{\rm X}/L_{\rm Edd})-2.29 \log P({\rm h})$; for the inclination of $i=70^\circ$: $M_{K{\rm ,model,i=70}}=3.71-0.70 \log (L_{\rm X}/L_{\rm Edd})-2.32 \log P({\rm h})$. For both cases the rms scatter of the data points from the approximation is $\sim0.03$. Here we adopted Eddington luminosity for 1.4~M$_\odot$ neutron star $L_{\rm Edd}=2\times10^{38}$ erg/s.

Following the idea of \cite{paradijs94} we construct the quantity $\Sigma_K$ in such way that it is proportional to NIR luminosity $L_K$ of the binary (but not to its absolute magnitude $M_K= -2.5\log L_K+{\rm const}$).
Therefore, using the average values of the obtained best-fitting parameters we
get $\Sigma_K$ for the near-infrared $K$ band: $\Sigma_{K}=(L_{\rm X}/L_{\rm
Edd})^{0.29} P({\rm h})^{0.92}$. For such parametrization the absolute
brightness $M_K=-2.5 \log \Sigma_K+{\rm const}$. Note, that derived scaling is
quite similar to our simplest theoretical estimates above.

\begin{figure}
\includegraphics[height=\columnwidth,bb=28 170 570 710,clip]{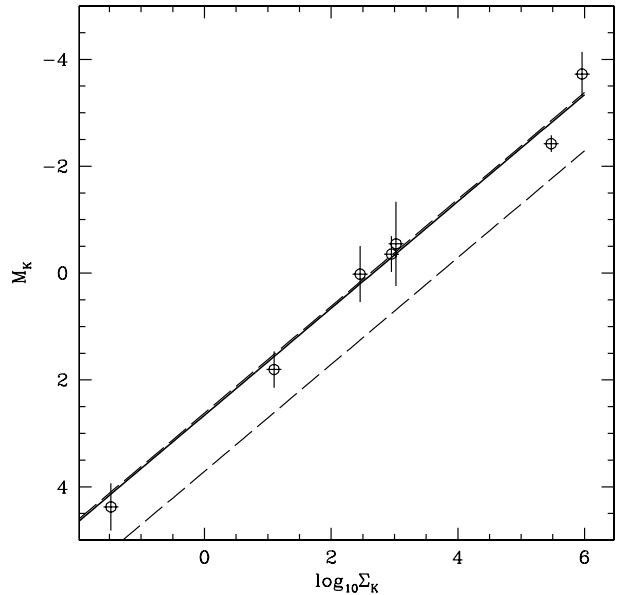}
\caption{Correlation between absolute NIR magnitude of persistent LMXBs and the combination of their X-ray luminosity with the orbital period, $\Sigma_{K}=(L_{\rm X}/L_{\rm Edd})^{0.29} P({\rm h})^{0.92}$. Crosses denote positions of known persistent LMXBs, dashed lines show best-fitting correlations calculated from our modelling, lower line -- for inclination $i=70^\circ$, upper line -- for $i=0^\circ$. Solid line shows the correlation $M_K=2.66-2.5\log \Sigma_K$}
\label{sigma_mk}
\end{figure}

\begin{table*}
\caption{Persistent LMXBs which were used to illustrate $\Sigma_{K}-M_{K}$ correlation plotted in Fig.~\ref{sigma_mk}. Literature references are given where necessary. X-ray luminosity $L_{\rm X}$ is taken from \citet{revnivtsev11} who calculated it using the X-ray flux from {\it Uhuru} catalog \citet{forman78} and the distance estimates available from the literature; orbital periods $P$ are adopted from \citet{ritter03}. Values of $A_K$ for 3 sources with detailed SED analysis in this paper were calculated from  $A_V$ values assuming $A_K=0.11A_V$. Values of $A_K$ for remaining sources were taken from 3D Galaxy extinction map by \citet{marshall06} if not indicated another reference. Magnitude of 4U0614+091 in the $K$ band was recalculated from $J$ measurement of \citet{migliari10} taking into account $J-K$ colours from the discussed model (see the text).}
\label{tab_K_and_P}
\begin{threeparttable}
\begin{tabular}{|l|c|c|c|c|r|r|r|l|l|r|}
\hline
  \multicolumn{1}{|l|}{Name} &
  \multicolumn{1}{c|}{$\log L_{\rm X}$ (2--10 keV)} &
  \multicolumn{1}{c|}{$d$} &
  \multicolumn{1}{c|}{$P$} &
  \multicolumn{1}{c|}{$m_{K}^{\rm corr}$} &
  \multicolumn{1}{c|}{$A_K$} \\

  \multicolumn{1}{|c|}{} &
  \multicolumn{1}{c|}{erg~s$^{-1}$} &
  \multicolumn{1}{c|}{kpc} &
  \multicolumn{1}{c|}{h} &
  \multicolumn{1}{c|}{Vega mag} &
  \multicolumn{1}{c|}{mag} \\

\hline
  Sco X-1 & 38.3 & 2.8 $\pm$ 0.3\tnote{1} & 18.94 &  11.88 & 0.1\\
  Cyg X-2 & 38.3 & 11.6 $\pm$ 0.3 \tnote{2} & 236.27 & 12.9 & 0.15\\
  GX 349+2 & 38.2 & 8.5\tnote{3} & 22.5 & 14.1\tnote{4} & 0.45\\
  GX 13+1 & 37.7 & 7 $\pm$ 1\tnote{5} & 601.7 & 10.5\tnote{6} & 1.8\tnote{7}\\
  4U1624-49 & 37.5 & 15.0 $\pm$ 2.9\tnote{8} & 20.9 & 15.9\tnote{9} & 2.4\\
  4U1735-44 & 37.7 &9.1	\tnote{10}	&4.65	& 16.6\tnote{89}&0.16\\
  4U1636-53 & 37.4 &5.9  &  3.79 & 15.9\tnote{11} &0.27\tnote{11}\\
  4U0614+091& 36.5 & 3.2&0.81&	16.9&0.22\\
\hline
\end{tabular}
\begin{tablenotes}
\item (1) -- \cite{bradshaw99}, (2) -- \cite{smale98}, (3) -- Galactic Center distance, (4) -- \cite{wachter96}, (5) -- \cite{bdp99}, (6) -- \cite{bdp02}, (7) -- \cite{cn92}, (8) -- \cite{xiang07}, (9) -- \cite{wachter05}, (10) -- \cite{augusteijn98}, (11) -- \cite{russell11}
\end{tablenotes}
\end{threeparttable}
\end{table*}

\subsection{Sample of known LMXBs}

Having in hand physically motivated value of $\Sigma_{K}$, on which the LMXB luminosity should depend, we constructed $\log \Sigma_{K}-M_{K}$ relation for the list of known LMXBs with NIR brightness measurements.

We have compiled a sample of all known LMXBs with measured values of their brightness in $K$ band (see Table~\ref{tab_K_and_P}).
We have selected only binaries with neutron stars in order to reduce possible additional scatter due to compact object mass uncertainties.

The relationship between absolute magnitude in $K$ band and $\log \Sigma_{K}$ for known binaries from Table~\ref{tab_K_and_P} is shown in Fig.~\ref{sigma_mk}.

We have used linear least squares to fit this dependence without taking into account measurement uncertainties because of the possible physical dispersion of the absolute NIR magnitude values due to the unknown inclinations of binaries.

The best-fitting approximation of this relation for all considered binaries is
$M_{K}=(2.78\pm0.24)-(2.60\pm0.11)\log \Sigma_{K}$ ($1\sigma$ confidence intervals). In order to estimate the confidence intervals here (and below) we assumed the uncertainties of data points corresponding to the unity value of the resulted $\chi^2$/d.o.f. Confidence intervals then were calculated as usual from interval of parameter which gives $\Delta \chi^2=1$.

Note that if our parametrization of this relationship is correct, then the coefficient in front of $\log \Sigma_{K}$ should be equal to 2.5 (because the constructed value $\Sigma_{K}$ is proportional to the IR luminosity $L_K$ of the system, whereas absolute brightness is $M_K=-2.5\log L_K+{\rm const}$), which is compatible with the results of the fit.
Fixing this coefficient we obtain:

$$
M_{K}=(2.66\pm0.11)-2.5\log \Sigma_{K}
$$

RMS scatter of the observed points from the fitted straight line for our sample of sources is $\sim$0.3~mag. Note, that this value is approximately a lower limit for any $M_K=f(L_{\rm X}, P)$ model which does not take into account explicitly the binary system inclination and the time variability of its flux.

As a useful application we also present here the approximation of $V-K$ and $J-K$ colours, obtained with our model, as a function of $P({\rm h})$ and $L_{\rm X}$ for orbital plane inclination $i=0^\circ$. The best fit was calculated minimizing the root mean square residual on logarithmic scale:

$$
\log (V-K+1.12)= -0.27\log L_{\rm X}/L_{\rm Edd}+0.34\log P({\rm h})-0.75
$$

$$
\log (J-K+0.28) = -0.27 \log L_{\rm X}/L_{\rm Edd}+0.33 \log P({\rm h})-1.31
$$

\begin{figure}
\includegraphics[height=0.9\columnwidth,bb=28 170 570 710,clip]{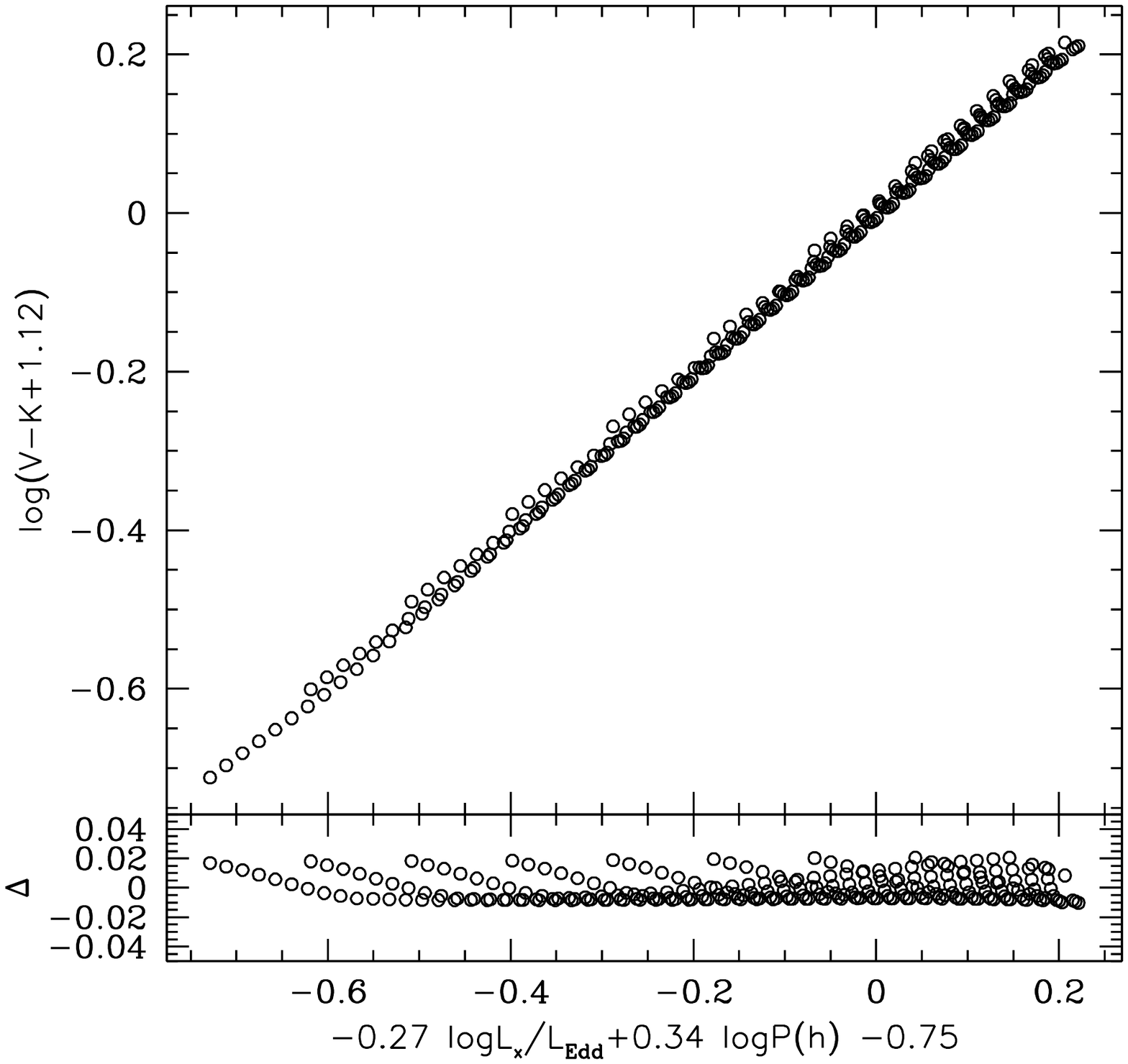}
\includegraphics[height=0.9\columnwidth,bb=28 170 570 710,clip]{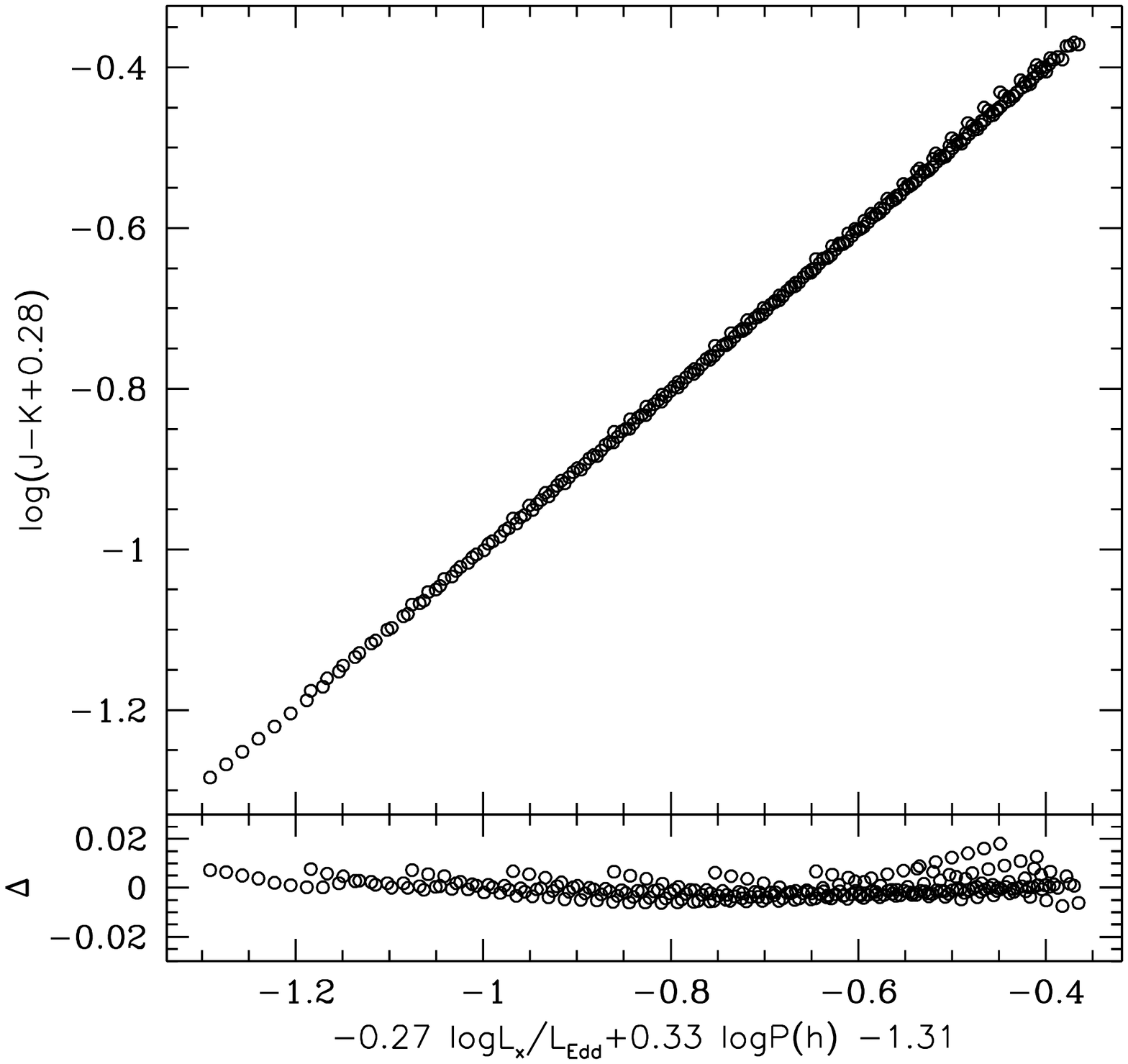}
\caption{Approximation of LMXB colors with models, described in the text. Upper plot shows $V-K$ color, lower plot -- $J-K$ color. Lower panels on each plot represent residuals from the adopted model.}
\label{colors}
\end{figure}

Residuals of these fits do not typically exceed 0.02--0.04 in logarithmic scale
(see Fig.~\ref{colors}).

\section{Summary}

In our paper we studied relations between the X-ray luminosity, orbital period, and absolute NIR magnitude of persistent low-mass X-ray binaries.

We have demonstrated that LMXB spectral energy distribution in the optical--NIR spectral range can often be adequately described by a simple model where all the surface elements in a binary system (comprising an accretion disc and a secondary star) emit black body radiation with the local temperature, which is itself defined by the illuminating X-ray flux. According to this model we can construct the quantity  $\Sigma_{K}=(L_{\rm X}/L_{\rm Edd})^{0.29} P({\rm h})^{0.92}$ on which the infrared luminosity of LMXBs depends almost exclusively. Therefore we were finally able to define the clear relation between $\Sigma_{K}$ and the absolute magnitude of LMXBs in the $K$ band $M_K$:

$$
M_{K}=(2.66\pm0.11)-2.5\log \Sigma_{K}
$$

Due to the fact that for the majority of persistent LMXBs we do have
estimates of their X-ray luminosities, the essence of this relation is that it
connects orbital period and absolute NIR magnitude of a binary system, hence one
may consider it to be effectively the period--luminosity relation in the NIR.

In some cases, however, we do see a significant emission excess in the NIR with the respect to the prediction of this model suggesting the presence of an additional spectral component (likely optically thin synchrotron emission from non-thermal electrons). We therefore recommend checking colours in the NIR spectral bands against expected values using the formula we give for an adequate usage of the presented $\Sigma_{K}-M_{K}$ relation. Extremely red colour indexes are not consistent with the colour of an optically thick regions in the binary systems considered in this work.

We propose that presented period--magnitude relation can be widely used in forthcoming surveys of the Galaxy in X-ray and NIR spectral domains.

\section*{Acknowledgements}
This research has made use of the VizieR catalogue access tool, CDS, Strasbourg,
France and NASA/ IPAC Infrared Science Archive, which is operated by the Jet Propulsion Laboratory, California Institute of Technology, under contract with the National Aeronautics and Space Administration. IYZ and AVM were supported by Russian state contract No.~02.740.11.0575,
MGR and AVM acknowledge the support by the grants of President of Russian Federation MD-1832.2011.2, RFBR 10-02-00492, 12-02-00186, program P19 of Presidium of the Russian Academy of Sciences/RAS and program OFN17 of the Division of Physical Sciences of the RAS. MGR is supported by Dynasty foundation.

\label{lastpage}


\begin{thebibliography}{}
\bibitem[\protect\citeauthoryear{Augusteijn et 
al.}{1998}]{augusteijn98} Augusteijn T., van der Hooft F., de Jong J.~A., van Kerkwijk M.~H., van Paradijs J., 1998, A\&A, 332, 561 
\bibitem[\protect\citeauthoryear{Bandyopadhyay et 
al.}{1999}]{bdp99} Bandyopadhyay R.~M., Shahbaz T., Charles 
P.~A., Naylor T., 1999, MNRAS, 306, 417
\bibitem[\protect\citeauthoryear{Bandyopadhyay et 
al.}{2002}]{bdp02} Bandyopadhyay R.~M., Charles P.~A., 
Shahbaz T., Wagner R.~M., 2002, ApJ, 570, 793 
\bibitem[\protect\citeauthoryear{Bradshaw, Fomalont,
\& Geldzahler}{1999}]{bradshaw99} Bradshaw C.~F., Fomalont E.~B., Geldzahler B.~J., 1999, ApJ, 512, L121
\bibitem[\protect\citeauthoryear{Casares et 
al.}{2010}]{casares10} Casares J., Gonz{\'a}lez Hern{\'a}ndez 
J.~I., Israelian G., Rebolo R., 2010, MNRAS, 401, 2517
\bibitem[\protect\citeauthoryear{Charles 
\& Naylor}{1992}]{cn92} Charles P.~A., Naylor T., 1992, MNRAS, 255, 6P\bibitem[\protect\citeauthoryear{Charles
\& Coe}{2006}]{charles} Charles P.~A., Coe M.~J., 2006, In: Compact stellar X-ray sources. Edited by Walter Lewin \& Michiel van der Klis. Cambridge Astrophysics Series, No. 39. Cambridge, UK: Cambridge University Press, p.215
\bibitem[\protect\citeauthoryear{Corbel
\& Fender}{2002}]{corbel02} Corbel S., Fender R.~P., 2002, ApJ, 573, L35
\bibitem[\protect\citeauthoryear{Cutri et al.}{2011}]{cutri11} Cutri R.~M., et al., 2011, Explanatory Supplement to the WISE Preliminary Data Release Products  http://wise2.ipac.caltech.edu/ docs/ release/ prelim/ expsup/ wise\_prelrel\_toc.html
\bibitem[\protect\citeauthoryear{D'Amico et
al.}{2001}]{damico01} D'Amico F., Heindl W.~A., Rothschild
R.~E., Gruber D.~E., 2001, ApJ, 547, L147
\bibitem[\protect\citeauthoryear{Dubus, Hameury, 
\& Lasota}{2001}]{dubus01} Dubus G., Hameury J.-M., Lasota J.-P., 2001, A\&A, 373, 251 \bibitem[\protect\citeauthoryear{de Jong, van Paradijs, 
\& Augusteijn}{1996}]{dejong96} de Jong J.~A., van Paradijs J., Augusteijn T., 1996, A\&A, 314, 484 
\bibitem[\protect\citeauthoryear{Evans et al.}{2010}]{evans10}
Evans I.~N., et al., 2010, ApJS, 189, 37
\bibitem[\protect\citeauthoryear{Fabbiano}{2006}]{fabbiano06} Fabbiano G., 2006, ARA\&A, 44, 323 
\bibitem[\protect\citeauthoryear{Fazio et al.}{2004}]{fazio04}
Fazio G.~G., et al., 2004, ApJS, 154, 10
\bibitem[\protect\citeauthoryear{Fender et al.}{1999}]{fender99}
Fender R.~P., Garrington S.~T., McKay D.~J., Muxlow T.~W.~B., Pooley G.~G.,
Spencer R.~E., Stirling A.~M., Waltman E.~B., 1999, MNRAS, 304, 865
\bibitem[\protect\citeauthoryear{Forman et al.}{1978}]{forman78} 
Forman W., Jones C., Cominsky L., Julien P., Murray S., Peters G., 
Tananbaum H., Giacconi R., 1978, ApJS, 38, 357 
\bibitem[\protect\citeauthoryear{Giacconi et
al.}{1962}]{giacconi62} Giacconi R., Gursky H., Paolini F.~R.,
Rossi B.~B., 1962, PhRvL, 9, 439
\bibitem[\protect\citeauthoryear{Hakala, Charles, 
\& Muhli}{2011}]{hakala11} Hakala P.~J., Charles P.~A., Muhli P., 2011, MNRAS, 416, 644 
\bibitem[\protect\citeauthoryear{Jimenez-Garate, Raymond,
\& Liedahl}{2002}]{jimenez02} Jimenez-Garate M.~A., Raymond J.~C., Liedahl D.~A., 2002, ApJ, 581, 1297
\bibitem[\protect\citeauthoryear{Kuulkers et 
al.}{2010}]{kuulkers10} Kuulkers E., et al., 2010, A\&A, 514, A65 
\bibitem[\protect\citeauthoryear{Lasota}{2001}]{lasota01} Lasota 
J.-P., 2001, NewAR, 45, 449 
\bibitem[\protect\citeauthoryear{Liu, van Paradijs,
\& van den Heuvel}{2007}]{liu07} Liu Q.~Z., van Paradijs J., van den Heuvel E.~P.~J., 2007, A\&A, 469, 807
\bibitem[\protect\citeauthoryear{London, McCray, 
\& Auer}{1981}]{london81} London R., McCray R., Auer L.~H., 1981, ApJ, 243, 970 
\bibitem[\protect\citeauthoryear{Marshall et 
al.}{2006}]{marshall06} Marshall D.~J., Robin A.~C., Reyl{\'e} C., Schultheis M., Picaud S., 2006, A\&A, 453, 635 
\bibitem[\protect\citeauthoryear{McClintock et
al.}{1979}]{mcclintock79} McClintock J.~E., Canizares C.~R.,
Cominsky L., Li F.~K., Lewin W.~H.~G., van Paradijs J., Grindlay J.~E.,
1979, Natur, 279, 47
\bibitem[\protect\citeauthoryear{McClintock et 
al.}{1984}]{mcclintock84} McClintock J.~E., Remillard R.~A., Petro 
L.~D., Hammerschlag-Hensberge G., Proffitt C.~R., 1984, ApJ, 283, 794 
\bibitem[\protect\citeauthoryear{McNamara et
al.}{2003}]{mcnamara03} McNamara B.~J., et al., 2003, AJ, 125,
1437
\bibitem[\protect\citeauthoryear{McNamara et
al.}{2005}]{mcnamara05} McNamara B.~J., Norwood J., Harrison
T.~E., Holtzman J., Dukes R., Barker T., 2005, ApJ, 623, 1070 \bibitem[\protect\citeauthoryear{Mescheryakov, Shakura, 
\& Suleimanov}{2011}]{mescheryakov11a} Mescheryakov A.~V., Shakura N.~I., Suleimanov V.~F., 2011, AstL, 37, 311 
\bibitem[\protect\citeauthoryear{Mescheryakov, Revnivtsev, 
\& Filippova}{2011}]{mescheryakov11b} Mescheryakov A.~V., Revnivtsev M.G., Filippova E.V., 2011, AstL, 37, 892
\bibitem[\protect\citeauthoryear{Migliari et
al.}{2007}]{migliari07} Migliari S., et al., 2007, ApJ, 671, 706
\bibitem[\protect\citeauthoryear{Migliari et
al.}{2010}]{migliari10} Migliari S., et al., 2010, ApJ, 710, 117
\bibitem[\protect\citeauthoryear{O'Brien et
al.}{2002}]{obrien02} O'Brien K., Horne K., Hynes R.~I., Chen
W., Haswell C.~A., Still M.~D., 2002, MNRAS, 334, 426
\bibitem[\protect\citeauthoryear{O'Brien et 
al.}{2004}]{obrien04} O'Brien K., Horne K., Gomer R.~H., Oke 
J.~B., van der Klis M., 2004, MNRAS, 350, 587
\bibitem[\protect\citeauthoryear{Orosz
\& Kuulkers}{1999}]{orosz99} Orosz J.~A., Kuulkers E., 1999, MNRAS, 305, 132 \bibitem[\protect\citeauthoryear{Paczynski}{1977}]{p77} 
Paczynski B., 1977, ApJ, 216, 822 
\bibitem[\protect\citeauthoryear{Paizis et
al.}{2006}]{paizis06} Paizis A., et al., 2006, A\&A, 459, 187
\bibitem[\protect\citeauthoryear{Pandey et
al.}{2007}]{pandey07} Pandey M., Rao A.~P., Ishwara-Chandra C.~H., Durouchoux P., Manchanda R.~K., 2007, A\&A, 463, 567
\bibitem[\protect\citeauthoryear{Postnov 
\& Kuranov}{2005}]{postnov05} Postnov K.~A., Kuranov A.~G., 2005, AstL, 31, 7 
\bibitem[\protect\citeauthoryear{Revnivtsev et
al.}{2011}]{revnivtsev11} Revnivtsev M., Postnov K., Kuranov A., Ritter H., 2011, A\&A, 526, A94
\bibitem[\protect\citeauthoryear{Rieke
\& Lebofsky}{1985}]{rieke85} Rieke G.~H., Lebofsky M.~J., 1985, ApJ, 288, 618 \bibitem[\protect\citeauthoryear{Rieke et al.}{2004}]{rieke04}
Rieke G.~H., et al., 2004, ApJS, 154, 25
\bibitem[\protect\citeauthoryear{Ritter
\& Kolb}{2003}]{ritter03} Ritter H., Kolb U., 2003, A\&A, 404, 301 
\bibitem[\protect\citeauthoryear{Russell et
al.}{2006}]{russell06} Russell D.~M., Fender R.~P., Hynes R.~I.,
Brocksopp C., Homan J., Jonker P.~G., Buxton M.~M., 2006, MNRAS, 371, 1334
\bibitem[\protect\citeauthoryear{Russell, Fender,
\& Jonker}{2007}]{russell07} Russell D.~M., Fender R.~P., Jonker P.~G., 2007, MNRAS, 379, 1108
\bibitem[\protect\citeauthoryear{Russell
\& Fender}{2008}]{russell08} Russell D.~M., Fender R.~P., 2008, MNRAS, 387, 713
\bibitem[\protect\citeauthoryear{Russell et 
al.}{2011}]{russell11} Russell D.~M., O'Brien K., 
Mu{\~n}oz-Darias T., Casella P., Gandhi P., Revnivtsev M.~G., 2011, arXiv, 
arXiv:1109.1839 
\bibitem[\protect\citeauthoryear{Shahbaz et
al.}{2008}]{shahbaz08} Shahbaz T., Fender R.~P., Watson C.~A.,
O'Brien K., 2008, ApJ, 672, 510
\bibitem[\protect\citeauthoryear{Smale}{1998}]{smale98} Smale 
A.~P., 1998, ApJ, 498, L141 \bibitem[\protect\citeauthoryear{Steeghs 
\& Casares}{2002}]{steeghs02} Steeghs D., Casares J., 2002, ApJ, 568, 273 \bibitem[\protect\citeauthoryear{Tjemkes, van Paradijs, 
\& Zuiderwijk}{1986}]{tjemkes86} Tjemkes S.~A., van Paradijs J., Zuiderwijk E.~J., 1986, A\&A, 154, 77 
\bibitem[\protect\citeauthoryear{Tout et al.}{1996}]{tout96} 
Tout C.~A., Pols O.~R., Eggleton P.~P., Han Z., 1996, MNRAS, 281, 257 \bibitem[\protect\citeauthoryear{van Paradijs
\& McClintock}{1994}]{paradijs94} van Paradijs J., McClintock J.~E., 1994, A\&A, 290, 133
\bibitem[\protect\citeauthoryear{Vrtilek et
al.}{1990}]{vrtilek90} Vrtilek S.~D., Raymond J.~C., Garcia M.~R., Verbunt F., Hasinger G., Kurster M., 1990, A\&A, 235, 162
\bibitem[\protect\citeauthoryear{Vrtilek et 
al.}{1991}]{vrtilek91} Vrtilek S.~D., Penninx W., Raymond J.~C., 
Verbunt F., Hertz P., Wood K., Lewin W.~H.~G., Mitsuda K., 1991, ApJ, 376, 
278 
\bibitem[\protect\citeauthoryear{Wachter 
\& Margon}{1996}]{wachter96} Wachter S., Margon B., 1996, AJ, 112, 2684 \bibitem[\protect\citeauthoryear{Wachter et 
al.}{2005}]{wachter05} Wachter S., Wellhouse J.~W., Patel S.~K., 
Smale A.~P., Alves J.~F., Bouchet P., 2005, ApJ, 621, 393 
\bibitem[\protect\citeauthoryear{Werner et 
al.}{2006}]{werner06} Werner K., Nagel T., Rauch T., Hammer N.~J., Dreizler S., 2006, A\&A, 450, 725 
\bibitem[\protect\citeauthoryear{Willis et al.}{1980}]{willis80}
Willis A.~J., et al., 1980, ApJ, 237, 596
\bibitem[\protect\citeauthoryear{Wright et al.}{2010}]{wright10}
Wright E.~L., et al., 2010, AJ, 140, 1868
\bibitem[\protect\citeauthoryear{Xiang, Lee, 
\& Nowak}{2007}]{xiang07} Xiang J., Lee J.~C., Nowak M.~A., 2007, ApJ, 660, 1309
\end{thebibliography}
\end{document}